\documentclass[aps,prd,superscriptaddress,nofootinbib,eqsecnum,twocolumn]{revtex4-1}

\pdfoutput=1

\usepackage{graphicx}  
\usepackage{dcolumn}   
\usepackage{amssymb}   
\usepackage{xcolor}
\usepackage{graphics} 
\usepackage{hyperref}
\usepackage{amsfonts}
\usepackage{latexsym}
\usepackage{slashed}
\usepackage{amsmath}
\usepackage{amsfonts}
\usepackage{amssymb}
\usepackage{dsfont}
\usepackage{bm}  
\usepackage[normalem]{ulem} 
\usepackage{amsmath,mathtools,mathrsfs,bbm,amssymb} 
\usepackage{bbold}
\usepackage{subfigure}
\usepackage{array}
\usepackage{wrapfig}
\usepackage{multirow}
\usepackage{tabularx}
\usepackage{float}
\usepackage{array}
\usepackage{booktabs}
\usepackage{commath}
\usepackage{graphicx}
\usepackage{epstopdf}
\usepackage{amsmath}
\usepackage{amsfonts}
\usepackage{amssymb}
\usepackage{dsfont}
\usepackage{latexsym}
\usepackage{hyperref}
\usepackage[english]{babel}
\usepackage[utf8]{inputenc}
\usepackage[colorinlistoftodos]{todonotes}
\usepackage{xcolor}
\usepackage{slashed}
\usepackage{bm}  
\usepackage[normalem]{ulem} 
\usepackage{amsmath,mathtools}
\usepackage{caption}
\usepackage{subcaption}
\usepackage{float}
\usepackage{MnSymbol}
\useunder{\uline}{\ul}{}
\usepackage{hyperref}
\usepackage{tikz}
\usepackage[compat=1.1.0]{tikz-feynman}

\begin{document}

\title{Non-Hermitian Dirac theory from Lindbladian dynamics}

\author{Y. M. P. Gomes} \email{yurimullergomes@gmail.com}
\affiliation{Departamento de F\'{\i}sica Te\'{o}rica, Universidade do Estado do Rio de Janeiro, Rio de Janeiro, RJ 20550-013, Brazil}

\begin{abstract} 

This study investigates the intricate relationship between dissipative processes of open quantum systems and the non-Hermitian quantum field theory of relativistic fermionic systems. By examining the influence of dissipative effects on Dirac fermions via Lindblad formalism, we elucidate the effects of coupling relativistic Dirac particles with the environment and show the lack of manifest Lorentz invariance. Employing rigorous theoretical analysis, we generalize the collisionless Boltzmann equations for the relativistic dissipation-driven fermionic system and find the Lyapunov equation, which governs the stationary solutions. Using our formalism, one presents a simple non-Hermitian model that the relativistic fermionic particles and anti-particles are stable. Going further, using the solution to the Lyapunov equations, one analyses the effect of dissipation on the stationary charge imbalance of this non-Hermitian model and finds that the dissipation can induce a new kind of charge imbalance compared with the collisionless equilibrium case.
\end{abstract}

\maketitle



\section{Introduction}
The theoretical description of the subatomic realm through quantum field theory (QFT) evolved significantly over the past century since Dirac's pioneering work in quantizing the electromagnetic field \cite{Dirac27}. The success of the Hamiltonian paradigm and its philosophical implications became hegemonic in the theoretical studies on QFT. As a consequence, the development of mathematical tools for describing the thermodynamic properties of quantum systems, based on Matsubara's seminal contributions \cite{Matsubara55} utilizing the imaginary time formalism (known as Matsubara formalism), further revolutionized our understanding of quantum and cosmological realms. Based on this success, many physicists embraced the equilibrium framework implicit in the Matsubara formalism \cite{Abrikosov63}. Still, a parallel line of research emerged during this period, focusing on studying out-of-equilibrium QFT. Prominent figures in this pursuit include J. Schwinger \cite{Schwinger61}, L.P. Kadanoff \cite{Kadanoff62}, and L.V. Keldysh \cite{Kelsysh65}, who made significant contributions to the field of non-equilibrium QFT.

Notably, the equilibrium philosophical standpoint hinders a genuine, although incomplete, comprehension of the nature of the kinetic equation governing a specific system. However, recognizing the significance of non-equilibrium quantum phenomena is vital for a deeper understanding of physical phenomena and for uncovering essential features beyond the equilibrium framework, as can be seen in recent research areas such as relativistic transport phenomena \cite{kinref1}, time-crystals \cite{timecrystal1,timecrystal2}, Chiral magnetic and vortical effects in high-energy process \cite{Kharzeev2015}.

I. Prigogine, a Nobel Prize winner for discovering the so-called dissipative structures \cite{nprize}, identified a unique phenomenon exclusive to systems that are out of equilibrium and, therefore, in constant motion. This discovery brought significant advances in the understanding of nature, challenging the existing paradigm in physics by focusing on systems that are not static, which, according to Prigogine, systems in equilibrium are the exception, not the norm. 
According to him, quantum mechanics based on the Copenhagen interpretation could not explain irreversible phenomena since, when defined in a Hilbert space, it has its real eigenvalues and preserves time-reversal symmetry \cite{Petrosky}.

The mathematical framework that describes a similar quantum phenomenon has been shown independently from the seminal works written years before in parallel by V. Gorini, A. Kossakowski, and G. Sudarshan (GKS) \cite{Gorini},
and G. Lindblad \cite{Lindblad}. 
The so-called GKSL master equation, or Lindbladian, is one of the general forms of Markovian master equations describing open quantum systems. The Schrödinger equation is generalized for open quantum systems, e.g., a system weakly coupled to a Markovian reservoir. Within this framework, the system's dynamical behavior is no longer unitary but preserves trace and remains the positivity of the density matrix regardless of initial conditions.

To avoid imposing equilibrium conditions, this study adopts the out-of-equilibrium formalism based on the real-time approach, in contrast to the imaginary time method employed in Matsubara formalism. Specifically, we utilize the Closed-Time-Path (CTP) formalism, originally developed by Schwinger \cite{Martin59}  and Keldysh \cite{Keldysh65}. For a detailed understanding of the formalism, interested readers can refer to references such as  Berges et al. \cite{Berges04},  Kamenev \cite{kamenev1,kamenev2,kamenev3} and Sieberer \cite{Sieberer}. Generally, the applications of the Lindblad formalism describe non-relativistic systems \cite{Prosen,Buchhold,Blais,Diehl}, usually in the condensed matter area. Recently, the possibility to describe relativistic systems has been analyzed in the bosonic case \cite{Diosi,Matsumura}, and the result shows the dissipation character breaks the manifest Lorentz invariance. Nonetheless, the lack of studies of relativistic fermionic systems remains.

Furthermore, the Lindblad master equation usually results in non-Hermitian properties on the effective action. The non-Hermitian quantum field theory is a recent field of study, and some exciting features arose, such as the violation of the Noether theorem that related symmetries and charge conservation and new sources of phase transitions  \cite{PT1,PT2,PT3}, relativistic non-Hermitian quantum mechanics \cite{Jones-Smith:2009qeu},  $\mathcal{PT}$-symmetric quantum field theories \cite{Bender:2005hf},\cite{Bender:2004vn}. However, compared with other approaches the GKSL formalism brings the opportunity to shed light on the structural properties of the system {\it a priori}, as the unitarity, trace preservation, and positivity properties cited before.

This work is organized as follows: Section 2 introduces the main aspects of the Lindblad formalism and its application to the relativistic fermionic system with linear jump operators. Using the path integral formalism one reaches the action that describes the system. Using the Dyson-Schwinger (DS) equations, one finds a generalization of the collisionless Boltzmann equations and looks at the stationary configurations of the kinetic equation, where the Lyapunov equations arose. One also analyses the linear response due to an external electromagnetic field. 
In Section 3, we apply our results in a model with some physical assumptions, analyze the system's stability, and look for the model's electromagnetic response. In Section 4, we present our final comments and perspectives. This paper will consider the natural units where $\hbar = k_b=c=1$. One uses throughout the paper $[A,B] = A B - BA $ and $\{A,B\} = AB+ BA$, and the space-time to have $D=3+1$ dimensions.

\section{Lindbladian dynamics for relativistic fermions:}
 To study the quantum open system, one focuses on the reduced density matrix $\rho$ of the system of interest, where one traces over the environment degrees of freedom. The result is the effective non-equilibrium dynamics governed by the following equation:
\begin{eqnarray}\label{lind}
\partial_t \rho = \mathcal{L} \rho~~,
\end{eqnarray}    
where
\begin{equation}
\mathcal{L} \rho = - i [H,\rho ] +\sum_i \lambda_i \left( {L}_i^\dagger \rho {L}_i -\frac{1}{2}\{ {L}_i^\dagger L_i, \rho \} \right) ~~, 
\end{equation}
with $H$ a Hermitian operator which can be interpreted as an effective Hamiltonian operator, the operators $L_i^\dagger, L_i$ are the non-Hermitian jump operators and model the coupling with the environment in combination with the positive constants $\lambda_i$. Equation \eqref{lind} is called the Lindblad master equation, and when the jump operators vanish, one recovers the Von-Neumann master equation of the density matrix. The systems described by equation \eqref{lind} are commonly called {\it driven open many-body quantum systems} and are the most general open quantum systems with time-local character.

Now, assuming that the fermionic system can be described by fermionic creation/annihilation operators $\hat{a}_{{\bf p},s}$, $\hat{b}_{{\bf p},s}$ that obey $\hat{a}_{{\bf p},s}| 0 \rangle =\hat{b}_{{\bf p},s}| 0 \rangle = 0$. Particle/antiparticle states are obtained by the action of the creation operators $\hat{a}_{{\bf p},s}^\dagger$ and $\hat{b}_{{\bf p},s}^\dagger$ operators, the creation operator for particle and antiparticle, respectively. They respect the Grassmman algebra $\{\hat{a}_{{\bf p},s}, \hat{a}_{{\bf p'},s'}^\dagger\} = \{\hat{b}_{{\bf p},s}, \hat{b}_{{\bf p'},s'}^\dagger\} = \delta_{s s'} \delta_{\bf p p'}$ and $\{\hat{a}_{{\bf p},s}, \hat{a}_{{\bf p'},s'}\} = \{\hat{b}_{{\bf p},s}, \hat{b}_{{\bf p'},s'}\}  = 0$. 
The simplest fermionic CP-even Hamiltonian can be written as:
\begin{equation}\label{H0}
    H_f = \sum_{\bf p,s} \epsilon_{\bf p}\left(  \hat{a}_{{\bf p},s}^\dagger \hat{a}_{{\bf p},s} + \hat{b}_{{\bf p},s}^\dagger \hat{b}_{{\bf p},s} \right) ~~,
\end{equation}
where $\epsilon_{\bf p}= \sqrt{{\bf p}^2 + m^2} $ .   This Hamiltonian \eqref{H0} describes a fermionic harmonic oscillator. To describe the dynamics of the fermionic quantum system in the presence of an environment with a short memory time, one assumes the simplest jump operators as linear operators as follows: 
\begin{equation}\label{jop}
L_{1}= \hat{a}_{{\bf p},s}~~,~~ L_2 = \hat{a}_{{\bf p},s}^\dagger ~~,~~L_{3}= \hat{b}_{{\bf p},s}~~,~~ L_4 = \hat{b}_{{\bf p},s}^\dagger ~~.
\end{equation}
To simplify the notation, we will use $\hat{c}$ to denote the creation/annihilation operators, where $c=a,b$. The Lindblad equation can be expressed as follows:
\begin{eqnarray}\nonumber
&&\mathcal{L} \rho = - i [ {H}_f,\rho ] +\sum_{ \sigma } \left[  \lambda_{1,\sigma} \left( \hat{c}_{\sigma}^\dagger \rho \hat{c}_{\sigma} -\frac{1}{2}\{ \hat{c}_{\sigma}^\dagger \hat{c}_{\sigma}, \rho \} \right) \right.
+\\
&&\hspace{1.5cm}+ \left. \lambda_{2,\sigma} \left( \hat{c}_{\sigma} \rho \hat{c}_{\sigma}^\dagger -\frac{1}{2}\{ \hat{c}_{\sigma} \hat{c}_{\sigma}^\dagger, \rho \} \right) \right] ~~.
\end{eqnarray}
where $\sigma = \{{\bf p}, s,c\}$ and $\lambda_{i,\sigma}={ \lambda_{i,{\bf p}, c}} \geq 0$ with $i=1,2$ defines the rate of particle gain and loss of the system, respectively. The sum over $c$ means the sum over the particle and antiparticles.

To express the reduced density matrix in the form of a functional integral, one can construct coherent states $|\Psi_{\bf p,s}\rangle,|\Psi^c_{\bf p,s}\rangle$ which obey  $\hat{a}_{{\bf p}, s}|\Psi_{\bf p,s}\rangle= \Psi_{\bf p,s}|\Psi_{\bf p,s}\rangle$ and $\hat{b}_{{\bf p}, s}|\Psi^c_{\bf p,s}\rangle= \Psi^c_{\bf p,s}|\Psi^c_{\bf p,s}\rangle$ the coherent particle/antiparticle state, respectively. The elements $ \Psi_{\bf p,s}$ and $ \Psi^c_{\bf p,s}$ are Grassmann variables. By use of the Grassmann algebra, one can write:
\begin{equation}
    |\Psi_{\sigma}\rangle = e^{- \Psi_{\sigma}\hat{c}_{\sigma}^\dagger }|0\rangle ~~, ~~ \langle\Psi_{\sigma}| = \langle 0| e^{- \tilde{\Psi}_{\sigma}\hat{c}_{\sigma} }
\end{equation}
with $\tilde{\Psi}_\sigma$ completely unrelated with $\Psi_\sigma$ and where $| 0 \rangle$ is the vacuum state defined via the identities $\hat{c}_\sigma |0\rangle = 0$, for $c = a,b$. It can be seen that  $\langle \Psi_{\sigma} |\Psi_{\sigma'} \rangle =\delta_{ \bf p p'} \delta_{cc'} \delta_{s s'} e^{ \tilde{\Psi}_{\sigma}\Psi_{\sigma'}}$, which means the coherent states are not normalized. Therefore, to fix this feature, one needs to introduce the so-called resolution of unity in the coherent states basis, which can be achieved as follows:
\begin{equation}
    \hat{1} = \prod_\sigma \int D\Psi_\sigma e^{-\tilde{\Psi}_{\sigma}\Psi_{\sigma} }|\Psi_{\sigma}\rangle \langle \Psi_{\sigma}| ~~.
\end{equation}
with $D\Psi= d\tilde{\Psi}_\sigma d\Psi_{\sigma} $. Moreover, with these coherent states, one can write any normally ordered operator in terms of the field  $\Psi_{\sigma}$. For instance:
\begin{equation}
    \langle \Psi_{\sigma'} | \hat{c}^\dagger_{\sigma'} \hat{c}_{\sigma} | \Psi_{\sigma} \rangle = \delta_{\sigma \sigma'} \tilde{\Psi}_\sigma\Psi_{\sigma} e^{\tilde{\Psi}_\sigma\Psi_{\sigma}}~~.
\end{equation}
Thus, the trace of some observable $\mathcal{O}$ is given by:

\begin{equation}
    Tr[\mathcal{O}] = \prod_\sigma \int D\Psi_\sigma e^{-\tilde{\Psi}_\sigma\Psi_{\sigma}} \langle \Psi_{\sigma}| \mathcal{O} |- \Psi_{\sigma}\rangle ~~. 
\end{equation}

Now, with these properties, one can construct the functional integral version of the reduced density matrix. The solution of equation \eqref{lind} can be formally written as:
\begin{equation}
    \rho(t) = \mathbb{T} e^{\int_{t_0}^t dt \hat{\mathcal{L}}(t)}\rho(t_0)
\end{equation}
where $\mathbb{T}$ represents the time-ordering and $\hat{\mathcal{L}}(t)$ is the general time-dependent Lindbladian super-operator. By use of discretization of time in $N$ slices $\delta_t$, labeled as $t_j$, $j=0,...,N$, the reduced density matrix at time $t_{j+1}$ can be written as follows:
\begin{equation}
    \rho_{j+1} = e^{\delta_t \hat{\mathcal{L}}_j}\rho_j \approx \left(1+\delta_t \hat{\mathcal{L}}_j \right)\rho_j~~.
\end{equation}
Due to the structure of the super-operator, one needs to introduce two sets of coherent states basis, $| \Psi_{{\bf p},j}^+ \rangle$, $| \Psi_{{\bf p},j}^- \rangle$,  acting from the left and right sides of the Lindbladian super-operator, respectively. Therefore, the reduced density matrix super-operator at $t_j$ can be written as:
\begin{eqnarray}\nonumber
    \hat{\rho}_{j} &=& \prod_\sigma\int \mathcal{D}(\Psi_{\sigma,j}) e^{-(\tilde{\Psi}^+_{\sigma_j}\Psi^+_{\sigma,j})+(\tilde{\Psi}^-_{\sigma_j}\Psi^-_{\sigma,j})}  \\
    && \hspace{1.cm} \times | \Psi^+_{\sigma,j}\rangle\langle \Psi^+_{\sigma,j}| \hat{\rho_j} | \Psi^-_{\sigma,j}\rangle \langle \Psi^-_{\sigma,j}|~~,
\end{eqnarray}
with $\mathcal{D}(\Psi_{\sigma,j})= d\Psi^+_{\sigma,j} d\Psi^-_{\sigma,j} $. The Lindblad super-operator at $t_j$ is given by:
\begin{eqnarray}\nonumber
\hat{\mathcal{L}}_j\left[|\Psi^+_{\sigma,j}\rangle \langle \Psi^-_{\sigma,j}| \right] &=& - i H_{j }|\Psi^+_{\sigma,j}\rangle \langle \Psi^-_{\sigma,j}| + \\\nonumber
&&\hspace{-2.8cm} + i |\Psi^+_{\sigma,j}\rangle \langle \Psi^-_{\sigma,j}|H_{j }+\sum_{\sigma',i} \lambda_{\sigma,i} \left( {L}_{\sigma',i}^\dagger |\Psi^+_{\sigma,j}\rangle \langle \Psi^-_{\sigma,j}| {L}_{\sigma',i} +\right.\\\nonumber
&&\hspace{-2.8cm} -\frac{1}{2}\left( {L}_{\sigma',i}^\dagger L_{\sigma',i} |\Psi^+_{\sigma,j}\rangle \langle \Psi^-_{\sigma,j}| +|\Psi^+_{\sigma,j}\rangle \langle \Psi^-_{\sigma,j}| {L}_{\sigma',i}^\dagger L_{\sigma',i}\right)~~.\\
\end{eqnarray}
Therefore, the matrix elements of the reduced density matrix in the subsequent instant of time $t_{j+1}$ can be written as:
\begin{eqnarray}\nonumber
&&\langle \Psi^+_{\sigma,j+1}| \hat{\rho}_{j+1} |\Psi^-_{\sigma,j} \rangle = \langle \Psi^+_{\sigma,j+1}| \hat{\rho}_{\sigma, j} |\Psi^-_{\sigma,j} \rangle+\\
&&\hspace{1.cm}+\delta_t   \langle \Psi^+_{\sigma,j+1}|  \hat{\mathcal{L}}_j\left[|\Psi^+_{\sigma,j}\rangle \langle \Psi^-_{\sigma,j}| \right] |\Psi^-_{\sigma,j} \rangle ~.
\end{eqnarray}
where using the properties of normal and time ordering and also the properties of the coherent states given by $ \hat{c}_{\sigma}| \Psi_{\sigma,j} \rangle = \Psi_{\sigma,j}| \Psi_{\sigma,j} \rangle$ and $ \langle \Psi_{\sigma,j} |\hat{c}_{\sigma}^\dagger = \langle \Psi_{\sigma,j}| \Psi_{\sigma,j}$, one reaches:
\begin{eqnarray}\nonumber
   && \langle \Psi^+_{\sigma,j+1}|  \hat{\mathcal{L}}_j \left[ |\Psi^+_{\sigma,j}\rangle \langle \Psi^-_{\sigma,j}| \right] |\Psi^-_{\sigma,j} \rangle   = \\\nonumber
   && \langle \Psi^+_{\sigma, j+1} |\Psi^+_{\sigma,j} \rangle \langle \Psi^-_{\sigma,j} | \Psi^-_{\sigma, j+1} \rangle \\
   && \times \mathcal{L}\left( \tilde{\Psi}_{\sigma,j+1}^\pm,\tilde{\Psi}_{\sigma,j}^\pm,\Psi_{\sigma, j}^\pm,\Psi_{\sigma, j+1}^\pm \right)~~.
\end{eqnarray}
The function $\mathcal{L}$ depends on the form of the Hamiltonian and the jump operators. Thus, at first order on $\delta_t$ one can affirm that:
\begin{eqnarray}\label{step}\nonumber
   && \langle \Psi^+_{\sigma,j+1}| \hat{\rho}_{j+1} |\Psi^-_{\sigma,j+1} \rangle = \\
   && =\prod_\sigma\int \mathcal{D}(\Psi_{\sigma,j})e^{\delta_t \mathcal{K}\left(  \Psi\right) }\langle \Psi^+_{\sigma,j}| \hat{\rho}_{\sigma, j} |\Psi^-_{\sigma,j} \rangle~~,
\end{eqnarray}
with 
\begin{eqnarray}\nonumber
\mathcal{K}\left(  \Psi\right)&=&\mathcal{K}\left( \tilde{\Psi}_{\sigma,j+1},\tilde{\Psi}_{\sigma,j},\Psi_{\sigma,j+1},\Psi_{\sigma,j}\right) = \\\nonumber
&&\hspace{-1.cm} =\frac{\left( \tilde{\Psi}^+_{\sigma, j+1}-\tilde{\Psi}^+_{\sigma, j}\right)}{\delta_t}{\Psi}^+_\sigma+\tilde{\Psi}^-_\sigma\frac{\left( \Psi^-_{\sigma, j+1}-\Psi^-_{\sigma, j}\right)}{\delta_t} +\\
&&+\mathcal{L}\left( \tilde{\Psi}_{\sigma,j+1}^\pm,\tilde{\Psi}_{\sigma,j}^\pm,\Psi_{\sigma, j}^\pm,\Psi_{\sigma, j+1}^\pm \right)~~.
\end{eqnarray}
Finally, by iterating expression from eq. \eqref{step} from the initial time $t_0$ to a final $t_f$ in the limit $\delta_t \rightarrow 0$ one finds:
\begin{equation}
\rho = \frac{e^{i S_{f}\left(\tilde{\Psi},\Psi\right)}}{Z}~~,~~Z = \int \mathcal{D}\left(\tilde{\Psi}, \Psi\right) e^{i S_{f}(\tilde{\Psi},\Psi)} ~~,
\end{equation}
with
\begin{eqnarray}\label{effaction}\nonumber
    S_{f}(\tilde{\Psi},\Psi) &=& \sum_{\sigma}\int_{t_0}^{t_f} dt \Bigg[ \tilde{\Psi}_{\sigma}^+ i \partial_t \Psi_{\sigma}^+ - \tilde{\Psi}_{\sigma}^- i \partial_t\Psi_{\sigma}^- -\\ \nonumber&&\hspace{-1.cm} - H_f\left( \tilde{\Psi}_{\sigma}^+,\Psi_{\sigma}^+\right) + H_f\left( \tilde{\Psi}_{\sigma}^-,\Psi_{\sigma}^-\right) + \\\nonumber
    && \hspace{-1.cm}- i \sum_{\sigma,i} \lambda_{i,\sigma}\left( (L_{i,\sigma}^*)^- L_{i,\sigma}^+ -\frac{1}{2}(L_{i,\sigma}^*)^+ L_{i,\sigma}^+ +\right. \\
    &&\hspace{-.5cm} \left. -\frac{1}{2}(L_{i,\sigma}^*)^- L_{i,\sigma}^- \right) \Bigg]~~,
\end{eqnarray}
where $\Psi^\pm_{\sigma} = \Psi^\pm_{\sigma}(t)$, $\tilde{\Psi}^\pm_{\sigma} = \tilde{\Psi}^\pm_{\sigma}(t)$  and  $L_{i,\sigma}^\pm= L_i\left( \tilde{\Psi}_{\sigma}^\pm,\Psi_{\sigma}^\pm\right) $.  Using the Hamiltonian and jump operators defined in eq. \eqref{H0} and eq. \eqref{jop}, one reaches: 

\begin{eqnarray}\nonumber
    &&S_{f}(\Psi^\pm) =\\\nonumber
    &&=\sum_{\sigma}\int_{t_0}^{t_f} dt \Bigg[ \tilde{\Psi}_{\sigma}^+( i \partial_t - \epsilon_{\sigma} )\Psi_{\sigma}^+ - \tilde{\Psi}_{\sigma}^-( i \partial_t - \epsilon_{\sigma} )\Psi_{\sigma}^- +\\\nonumber
    &&\hspace{1.cm} - i \lambda_{1,\sigma}\left(\tilde{\Psi}^-_{\sigma} \Psi^+_{\sigma} -\frac{1}{2}\tilde{\Psi}^+_{\sigma} \Psi^+_{\sigma}-\frac{1}{2}\tilde{\Psi}^-_{\sigma} \Psi^-_{\sigma} \right) +\\\nonumber
    &&\hspace{1.cm} + i \lambda_{2,\sigma}\left( \tilde{\Psi}^+_{\sigma} \Psi^-_{\sigma} -\frac{1}{2}\tilde{\Psi}^+_{\sigma} \Psi^+_{\sigma}-\frac{1}{2}\tilde{\Psi}^-_{\sigma} \Psi^-_{\sigma} \right) \Bigg] ~~.
\end{eqnarray}
Note the plus sign in front of $\lambda_{2,\sigma}$, which comes from the anti-Grassmann properties. Furthermore, one can introduce a new basis, the so-called Keldysh basis, given by:
\begin{equation}\label{kbasis}
    \Psi^{1/2}_\sigma = \frac{1}{\sqrt{2}}\left(\Psi^+_\sigma \pm \Psi^-_\sigma\right)~~,~~  \tilde{\Psi}^{1/2}_\sigma = \frac{1}{\sqrt{2}}\left(\tilde{\Psi}^+_\sigma \mp \tilde{\Psi}^-_\sigma\right)~~.
\end{equation}
So, the action can be rewritten as:
\begin{eqnarray}\nonumber\label{eq21}
    &&S_{f} = \sum_{\sigma}\int_{t_0}^{t_f} dt  \\\nonumber
    && \begin{pmatrix}
        \tilde{\Psi}^1_\sigma & \tilde{\Psi}^2_\sigma 
    \end{pmatrix} \begin{pmatrix}
        i \partial_t - \epsilon_\sigma - i q_\sigma & i d_\sigma \\
        0 & i \partial_t - \epsilon_\sigma + i q_\sigma
    \end{pmatrix} \begin{pmatrix}
        \Psi^1_\sigma \\
        \Psi^2_\sigma
    \end{pmatrix} ~~.\\
\end{eqnarray}

where $q_\sigma = \frac{1}{2} \left( \lambda_{1,\sigma} + \lambda_{2,\sigma} \right)$ and $d_\sigma = \left( \lambda_{1,\sigma} -\lambda_{2,\sigma} \right)$. This result is already obtained in Ref. \cite{kamenev1}. Now, for relativistic fermions, one can improve Eq. \eqref{eq21} and, based on the assumptions that the dissipative corrections are weak, one can write the action in terms of the following four-component Dirac spinors:
\begin{eqnarray}\nonumber
    &&\psi^{a}({\bf x}, t) = \sum_{{\bf p},s}  (2 \epsilon_{\bf p})^{-1/2}\left( \Psi^{a}_{{\bf p},s}(t) u_{{\bf p}, s} e^{ i {\bf p} \cdot {\bf x}} +\right.\\
    &&\hspace{2.cm}\left. + (\tilde{\Psi}^c)^{a}_{{\bf p}, s}(t) v_{{\bf p},s} e^{- i {\bf p} \cdot {\bf x}} \right) ~~,
\end{eqnarray}
with $a = 1,2$ and $u_{{\bf p},s}$, $v_{{\bf p},s}$ are shown in Appendix \ref{appA}, and defining the Dirac conjugate $\bar{\psi} = \psi^\dagger \gamma^0$ one has:
\begin{eqnarray}\nonumber
   &&\bar{\psi}^a({\bf x}, t) = \sum_{{\bf p},s} (2 \epsilon_{\bf p})^{-1/2}\left( \tilde{\Psi}^a_{{\bf p},s}(t) \bar{u}_{{\bf p}, s} e^{- i {\bf p} \cdot {\bf x}} \right.+\\
   &&\hspace{2.cm}+\left. ({\Psi}^c)^a_{{\bf p}, s}(t) \bar{v}_{{\bf p},s} e^{ i {\bf p} \cdot {\bf x}} \right) ~~.
\end{eqnarray}

Therefore, the fermionic action takes the final form given by $S_f = S_{f,0} + S_{d}$, where:
\begin{eqnarray}\label{sbath2}\nonumber
S_{f,0}(\bar{\psi},\psi) &=& \\
&&\hspace{-2.3cm} = \int dt d{\bf x}  \begin{pmatrix}
\bar{\psi}^{1} ~\bar{\psi}^2 
\end{pmatrix} \begin{pmatrix}
i \slashed{\partial} - m  &  0\\
0 & i \slashed{\partial}-m
\end{pmatrix} \begin{pmatrix}
\psi^{1}\\
\psi^2 \end{pmatrix}  ,
\end{eqnarray}
with $\slashed{\partial}=\gamma^\mu \partial_\mu = (\gamma^0\partial_t - {\bf \gamma} \cdot \nabla)$. The dissipative term reads:

\begin{eqnarray}\nonumber
    S_{d} =  \int dt d{\bf x} d{\bf x}'\left( \bar{\psi}^a(t,{\bf x}) \check{\mathcal{D}}_{ab}({\bf x}- {\bf x'}) \psi^b(t,{\bf x'}) \right)\\ ~~,
\end{eqnarray}
where the dissipative kernel is written as follows:
\begin{equation}
    \check{\mathcal{D}}_{ab}({\bf x}- {\bf x'}) =  \begin{pmatrix}
        { +} i \check{Q}({\bf x}- {\bf x'}) &  i  \check{D}({\bf x}- {\bf x'})\\
        0 &   { -} i  \check{Q}({\bf x}- {\bf x'}) 
    \end{pmatrix} ~~,~~ 
\end{equation}
where its components read:
\begin{equation}\label{Qform}
    \check{Q}({\bf x}- {\bf x'}) = \sum_{ \bf p} e^{i {\bf p} \cdot ({\bf x}- {\bf x'})}\check{Q}({\bf p})
\end{equation}
and
\begin{equation}\label{Dform}
    \check{D}({\bf x}- {\bf x'}) = \sum_{\bf p} e^{i {\bf p} \cdot ({\bf x}- {\bf x'})}\check{D}({\bf p}) ~~.
\end{equation}
with
\begin{equation}\label{eq29}
 {  \check{Q}({\bf p}) = \frac{ \epsilon_{\bf p}}{2 m} \Big( \left(1+\frac{\gamma_j{\bf p}_j}{m}\right)\left(q_{\bf p} - \tilde{q}_{\bf p}\right) + \frac{\epsilon_{\bf p}}{m} \left(q_{\bf p} + \tilde{q}_{\bf p}\right) \Big) ~~,}
\end{equation}
and
\begin{equation}\label{eq30}
 {\check{D}({\bf p}) = \frac{ \epsilon_{\bf p}}{2 m} \Big( \left(1+\frac{\gamma_j{\bf p}_j}{m}\right)\left(d_{\bf p} - \tilde{d}_{\bf p}\right) + \frac{\epsilon_{\bf p}}{m} \left(d_{\bf p} + \tilde{d}_{\bf p}\right) \Big) ~~,}
\end{equation}
and   {redefining}
\begin{equation}
\lambda_{i,{\bf p},c=a}=\lambda_{i,{\bf p}} ~~,~~\lambda_{i,{\bf p},c=b}=\tilde{\lambda}_{i,{\bf p}}~~,
\end{equation}
  {the dissipative couplings for the particles and antiparticles, respectively, one finds that $q_{\bf p} =\frac{1}{2} \left( \lambda_{1,{\bf p}} + \lambda_{2,{\bf p}} \right)$ and $d_{\bf p} = \left( \lambda_{1,{\bf p}} -\lambda_{2,{\bf p}} \right)$, $\tilde{q}_{\bf p} = \frac{1}{2} \left( \tilde{\lambda}_{1,{\bf p}} + \tilde{\lambda}_{2,{\bf p}} \right)$ and $\tilde{d}_{\bf p} = \left( \tilde{\lambda}_{1,{\bf p}} -\tilde{\lambda}_{2,{\bf p}} \right)$. Eqs. \eqref{eq29} and \eqref{eq30} are important new results in this work that describe the dissipative modifications on a relativistic fermionic system. The fact that the on-shell momentum appears in equations \eqref{Qform} and \eqref{Dform} is due to the on-shell profile of the particle/antiparticle projectors (defined in Appendix \ref{appA}). Thus, one shows that the relativistic fermionic system acquires a non-local behavior when interacting with the environment. Particularly, Lindbladian formalism provides a general form to describe the Markovian interaction process with the external bath via quantum jumps and generates a non-locality in space coordinates.}

\subsection{Dyson-Schwinger equation}
To construct the Green functions of the system, one defines the generating functional as follows:
\begin{equation}
Z[\bar{\eta}, \eta] = \int \mathcal{D}[\bar{\psi}, \psi] e^{i S[\bar{\psi}, \psi] + i\int dx \left( \bar{\eta}\psi + \bar{\psi}\eta \right) }~~. 
\end{equation}
with $\eta^a,\bar{\eta}^a$ the corresponding sources. The generating functional of the connected Green functions $W$ can be written as $W = \frac{1}{i} \ln Z$.
Thus, defining the effective action as $\Gamma[\bar{\psi},\psi] = W[\bar{\eta},\eta] - \int dx [\bar{\eta}\psi + \bar{\psi} \eta]$ one can find the Dyson-Schwinger (DS) equation for the full fermionic propagator $\check{\mathcal{S}}_{ab}(x,y)$ and it is given by:
\begin{equation}\label{DSeq}
\int dy\check{\mathcal{S}}^{-1}_{ab}(x,y)\check{\mathcal{S}}_{bc}(y,z) = \delta_{ac} \delta(x-z),
\end{equation}
with
\begin{equation}\label{Sdecomp}
\check{\mathcal{S}}^{-1}_{ab}(x,y) = \begin{pmatrix}
(\check{\mathcal{S}}^R)^{-1}(x,y) &  \check{\Delta}(x,y)\\
0 & (\check{\mathcal{S}}^A)^{-1}(x,y)
\end{pmatrix}~~,
\end{equation}
where its components read:
\begin{eqnarray}\nonumber
    &&(\check{\mathcal{S}}^{R/A})^{-1}(x,x') = \\\nonumber
    &&=\left( \left( i\slashed{\partial}- m \right)\delta({\bf x}- {\bf x}') { \pm} i \check{Q}({\bf x}- {\bf x}')\right)\delta(t-t')~~.\\
\end{eqnarray}
and
\begin{eqnarray}\nonumber
    &&\check{\Delta}(x,x') = i \check{D}({\bf x}- {\bf x}') \delta(t-t')~~.\\
\end{eqnarray}
Therefore, the fermionic propagator can be written compactly as follows:
\begin{equation}
\check{\mathcal{S}}_{ab}(x,y) = \begin{pmatrix}
\check{\mathcal{S}}^{R}(x,y) &  \check{\mathcal{S}}^K(x,y)\\
0 & \check{\mathcal{S}}^A(x,y)
\end{pmatrix}~~.
\end{equation}
In this way, the components of the DS equation can be expressed as follows:

\begin{equation}\label{DSeq2}
\int dy\left( \check{\mathcal{S}}^{R/A}\right)^{-1}(x,y)\check{\mathcal{S}}^{R/A}(y,z) =  \delta(x-z),
\end{equation}
and 
\begin{eqnarray}\label{DSeq3}\nonumber
&&\check{\mathcal{S}}^K(w,z) =-\int dx dy  \check{\mathcal{S}}^{R}(w,x)  \check{\Delta}  (x,y)\check{\mathcal{S}}^A(y,z)~~.
\end{eqnarray}

The retarded/advanced components of the propagator can be written in momentum space as follows:
\begin{equation}
    \check{\mathcal{S}}^{R/A}(x,y) = \sum_p e^{i  p \cdot ( x-y)}\check{\mathcal{S}}^{R/A}(p)~~,
\end{equation}
  with  
\begin{equation} 
    \check{\mathcal{S}}^{R/A}(p) = \frac{1}{\slashed{p} - m {  \pm } i \check{Q}(\bf p)}~~. 
\end{equation}

One can notice that the manifest Lorentz invariance is lost due to dissipation. Going further, in general, is a common procedure to redefine the anti-Hermitian Keldysh Green function $\mathcal{S}^K$ by a Hermitian function $\check{F} = \overline{(\check{F})}$ as $S^K = S^R \circ \check{F} - \check{F} \circ S^A$, where $\check{F}$ is called Wigner distribution and the symbol ``$\circ$'' represents the convolution operator. By using this parametrization and after some algebraic manipulations on eq. \eqref{DSeq3}, one reaches:

 \begin{eqnarray}\label{DSeq4}\nonumber
&& \check{\Delta}(x,w)+\\\nonumber
&&+\int dy\left[  \check{F}(x,y)\left( \check{\mathcal{S}}^A\right)^{-1}(y,w) -(\check{\mathcal{S}}^{R})^{-1}(x,y)\check{F}(y,w) \right] =  0.\\
\end{eqnarray}

Notice that eq.\eqref{DSeq4} is the kinetic equation of the system in coordinate space. Now, for a given two-point function $A(x,x')$, one can introduce new
variables designated by the central point coordinate $X = (x+x')/2$ and the relative coordinate $\xi = x-x'$, 
 one can perform the Fourier transform on the relative coordinate $\xi$ such that $A(x,x') \rightarrow A(X,p)$. This procedure is known as the Wigner transformation \cite{kamenev1}. Applying this technique the Eq. \eqref{DSeq4} one finds that the kinetic equation reads:

\begin{eqnarray}\nonumber\label{kineq}
&&\frac{i}{2}\left\{ \gamma^\mu, \frac{\partial \check{F}(X,p)}{\partial X^\mu} \right\}= [ \slashed{p}, \check{F}(X,p)] +\\
&&\hspace{1.cm}- i \{  \check{Q}(p) , \ostar \check{F}(X,p) \}  + i  \check{D}(p),
\end{eqnarray}

where 
\begin{eqnarray}\nonumber 
    &&A(X,p) \ostar B(X,p) = \\
    &&A(X,p) e^{\frac{i}{2}\overleftarrow{\partial}_X \overrightarrow{\partial}_p - \overleftarrow{\partial}_p \overrightarrow{\partial}_X } B(X,p)~~.
\end{eqnarray}
and the operation with $\ostar$ is known as the Moyal product. It is important to highlight that eq. \eqref{kineq} is a generalization of the Boltzmann equation for a relativistic open fermionic system and is an important result of this work. Using the Wigner transformation, the Keldysh component of the fermionic propagator reads:
\begin{equation}
    \check{\mathcal{S}}^K(X,p) = \check{\mathcal{S}}^R(p) \ostar \check{F}(X,p) - \check{F}(X,p) \ostar \check{\mathcal{S}}^A(p) ~~.
\end{equation}
Now, assuming a relatively slow dependence of the central coordinates $X_\mu$ one can rewrite Eq. \eqref{kineq} as follows: 
\begin{eqnarray}\nonumber\label{kineqexp}
&&\frac{i}{2}\left\{ \gamma^\mu, \frac{\partial \check{F}(X,p)}{\partial X^\mu} \right\} = [ \slashed{p}, \check{F}(X,p)] - i \{  \check{Q}(p) , \check{F}(X,p) \}   +  \\
&&+i  \check{D}(p)  -  \frac{1}{2} \left\{  \frac{\partial\check{Q}(p) }{\partial p_\mu} , \frac{\partial\check{F}(X,p)}{\partial X^\mu} \right\} + O(\partial_X^2)~~.
\end{eqnarray}

Now, it is useful to assume $\check{F}(X,p)$  independent of $X$ ($\check{F}(X,p)=\check{f}(p)$) and therefore a stationary solution, and apply it to eq. \eqref{kineq} one finds:
\begin{equation}\label{stkin}
    [ \slashed{p}, \check{f}(p)] - i \{  \check{Q}(p), \check{f}(p) \}  + i  \check{D}(p) = 0~~.
\end{equation}
 Therefore, one can affirm that Eq. \eqref{stkin} is the Lyapunov equation of the relativistic dissipation-driven fermionic system and is one of our main results.  To extract information from eq. \eqref{stkin}, based on Ref. \cite{kinref1}, one decomposes the stationary Wigner distribution $f(p)$ in terms of its spinorial structure as follows:
\begin{equation}
\check{f}(p)= \mathfrak{s}({p}) + i \gamma_5 \mathfrak{p}(p)+ \slashed{\mathfrak{v}}({p}) + i \gamma_5 \slashed{\mathfrak{a}}({p}) +\frac{1}{2} \sigma_{\mu \nu}\mathfrak{t}^{\mu \nu}({p})~~,
\end{equation}
 and the non-Hermitian terms can be decomposed as follows:
 \begin{equation}
     \check{Q}(p) = \slashed{\xi}^q(p) + \rho^q(p) ~~,~~ \check{D}(p) = \slashed{\xi}^d(p) + \rho^d(p) ~~.
 \end{equation}
  {with $\xi^q_0 = \frac{\epsilon_{\bf p}^2}{2 m^2} (q_{\bf p} + \tilde{q}_{\bf p})$, $\xi^q_i = \frac{\epsilon_{\bf p}}{2 m^2} {\bf p}_i (q_{\bf p} - \tilde{q}_{\bf p})$, $\xi^d_0 = \frac{\epsilon_{\bf p}^2}{2 m^2} (d_{\bf p} + \tilde{d}_{\bf p})$, $\xi^d_i = \frac{\epsilon_{\bf p}}{2 m^2} {\bf p}_i (d_{\bf p} - \tilde{d}_{\bf p})$}, $\rho^q = \frac{ \epsilon_{\bf p}}{2 m}  (q_{\bf p} - \tilde{q}_{\bf p})$ and $\rho^d = \frac{ \epsilon_{\bf p}}{2 m}  (d_{\bf p} - \tilde{d}_{\bf p})$. Combining these features on \eqref{stkin}, one finds the following set of equations:
\begin{subequations}
    \begin{equation}
    2 \mathfrak{s} \rho^q +\rho^d+2 \mathfrak{v} \cdot \xi^q = 0 ~~,
    \end{equation}
    \begin{equation}
    \mathfrak{a} \cdot p +\mathfrak{p} \rho^q = 0~~,
    \end{equation}
    \begin{equation}
    2 i p^\mu\mathfrak{t}_{\beta\mu}+\xi^d_{\beta }+2 \mathfrak{s} \xi^q_{\beta }+2 \rho^q \mathfrak{v}_{\beta } = 0~~,
    \end{equation}
    \begin{equation}
    -i  (\xi^q)^\mu ~^* \mathfrak{t}_{\beta \mu}-\rho^q \mathfrak{a}_{\beta }+\mathfrak{p} p_{\beta } = 0~~,
    \end{equation}
    \begin{equation}
    -i  \rho^q \mathfrak{t}_{\beta \rho } + p_{[\rho } \mathfrak{v}_{\beta ]} +\varepsilon _{\beta \rho \mu \nu} \mathfrak{a}^\mu (\xi^q)^\nu = 0~~.
    \end{equation}
\end{subequations}
with $^*\mathfrak{t}_{\beta \mu} = \frac{1}{2} \mathfrak{t}^{\nu \rho} \varepsilon _{\beta \mu \nu \rho}$. Despite the intricate character of those equations, the solution for each component of the Wigner distribution can be found and is given by:
\begin{subequations}
    \begin{eqnarray}\label{Lyapsol1}\nonumber
       \mathfrak{s} &=&\frac{-n_{\bf p}\left(\left(  p \cdot \xi^q\right)^2+6 (\rho^q)^2 (\xi^q)^2\right) }{2 \left((\rho^q)^2 \left({p}^2-6 (\xi^q)^2\right)+6 (\rho^q) ^4-\left(\xi^q \cdot {p}\right)^2\right)}+\\\nonumber
&&+\frac{m_{\bf p} (\rho ^q)^2\left( p^2+6  (\rho^q)^2\right) }{2 \left((\rho^q)^2 \left({p}^2-6 (\xi^q)^2\right)+6 (\rho^q) ^4-\left(\xi^q \cdot {p}\right)^2\right)}  ~,\\
    \end{eqnarray}
    \begin{eqnarray}\label{Lyapsol2}\nonumber
        &&\mathfrak{v}_\mu = \frac{\rho^q (n_{\bf p}-m_{\bf p}) \left(6 (\rho^q)^2 \xi^q_{\mu }+ \left(p \cdot  \xi^q \right){p}_{\mu }\right)}{2 \left((\rho^q)^2 \left({p}^2-6 (\xi^q)^2\right)+6 (\rho^q) ^4-\left(\xi^q \cdot {p}\right)^2\right)}  ~~,\\
    \end{eqnarray}
    \begin{equation}\label{Lyapsol3}
        \mathfrak{t}_{\rho \beta} = \frac{i}{6 \rho^q} p_{[\rho}   \mathfrak{v}_{\beta]} ~~,
        \end{equation}
    \end{subequations}
with $\mathfrak{a} = \mathfrak{p}=0$,  and where one defines $n_{\bf p}= \frac{d_{\bf p} + \tilde{d}_{\bf p}}{q_{\bf p} + \tilde{q}_{\bf p}}$ and $m_{\bf p} = \frac{d_{\bf p} - \tilde{d}_{\bf p}}{q_{\bf p} - \tilde{q}_{\bf p}}$.  Going further, in the next section one analyses the linear response to an external electromagnetic field.

\subsection{Electromagnetic response:}

To analyze the behavior of the fermionic system under the presence of an external electromagnetic field, we have to couple the fermions to the Electromagnetic field through the covariant derivative operator $D_\mu = \partial_\mu - i A_\mu$, following the U(1) gauge symmetry principle. Nonetheless, in the out-of-equilibrium paradigm, one must introduce two fields $A_\mu^+, A_\mu^-$ which interact with the system from the left and right sides. Moreover, rewriting the EM field in terms of classical and quantum fields as follows:
\begin{equation}
    A_\mu^{cl} = \frac{1}{2}\left( A_\mu^+ +A_\mu^- \right) ~~,~~A_\mu^{q} = \frac{1}{2}\left( A_\mu^+ -A_\mu^- \right) ~~,
\end{equation}
one can write the interaction action \cite{kamenev1}:
\begin{eqnarray}\nonumber
   && S_{A} =  e \int d{ x}  \Bigg[ \begin{pmatrix}
\bar{\psi}^{1} ~\bar{\psi}^2 
\end{pmatrix} \begin{pmatrix}
 \slashed{A}^{cl}  &   \slashed{A}^{q}\\
 \slashed{A}^{q} &   \slashed{A}^{cl}
\end{pmatrix} \begin{pmatrix}
\psi^{1}\\
\psi^2 \end{pmatrix}  \Bigg] \\
&&\hspace{1.cm}=e \int d{ x}  \bar{\psi}^a \gamma^\mu \psi^b A^c_\mu (\hat{\sigma}_c)_{ab} ~~,
\end{eqnarray}
where $a,b,c = \{ cl,q\}$ and $\hat{\sigma}^{cl} = \begin{pmatrix}
    1 & 0 \\
    0 & 1
\end{pmatrix}$, $\hat{\sigma}^{q} = \begin{pmatrix}
    0 & 1 \\
    1 & 0
\end{pmatrix}$. The presence of the quantum source field violates causality, but they are non-physical and only serve as auxiliary fields to generate observables. The electronic current can be calculated by application of a functional derivative as follows:
\begin{equation}\label{current}
    j^c_\mu (x)= \frac{\delta ln Z[A]}{\delta A^\mu_c(x)}\Bigg|_{A=0} = i e Tr \Big[ \gamma_\mu \check{\mathcal{S}}^{ab}(x,x) \left( \hat{\sigma^c}\right)_{ab}   \Big]~.
\end{equation}

The physical current is coupled to the field's quantum component, and the action's causality is restored by taking the quantum field as zero. Thus, one finds:
\begin{equation}
    j^{cl}_\mu (x)= \frac{\delta ln Z[A]}{\delta A^\mu_q(x)}\Bigg|_{A=0} = i e Tr \big[ \gamma_\mu \check{\mathcal{S}}^{K}(x,x)   \big] ~~.
\end{equation}
and $j_\mu^q = 0$ insuring trace preservation. Thus, one now can look for stationary configurations, and by  use of the stationary solution of the Lyapunov equation one reaches:
\begin{eqnarray}\label{jst}\nonumber
    j^{st}_\mu &=&  2 e Im \int dp Tr\left[ \gamma_\mu \check{\mathcal{S}}^{R}(p) \check{f}(p)\right]  ~~.
\end{eqnarray}

Now, one will apply our results to a special model to construct a non-Hermitian model. 

\section{Application: a pedagogical non-Hermitian model}\label{sec3}

For instance, assuming the following special case where:
\begin{equation}\label{cases}
       {d}_{\bf p}= \chi_{\bf p} {q}_{\bf p}  \text{ and } \tilde{d}_{\bf p} = {\chi}_{\bf p} \tilde{q}_{\bf p} ~~. 
\end{equation}
Therefore, in this special case, one can verify that $n_{\bf p}= m_{\bf p}= {\chi}_{\bf p}$ and therefore the only nonvanishing component of the Wigner distribution is given by $\mathfrak{s}(p)$.  Precisely, the Lyapunov solutions are shown in Eqs. \eqref{Lyapsol1},\eqref{Lyapsol2}, \eqref{Lyapsol3} simplifies to 
\begin{equation}\label{specialcase}
     \mathfrak{s}(p)= {\chi}_{\bf p} ~~\text{, and}~~ \mathfrak{v}=\mathfrak{t} = 0 ~~. 
\end{equation}

Notice that, in this case, the solution of the Lyapunov equation is on-shell, i.e., depends only on the spatial component of momentum coordinate $\bf p$.   {Thus, one uses $\chi_{\bf p}=\mathfrak{s}_{\bf p}$ from now on.} To shed light on the physical meaning of the choices we made in this section, the assumptions made in case 1 imply that the coupling constants can be rewritten as follows: 
\begin{eqnarray}
  \lambda_{1,{\bf p}} = q_{\bf p} \left( 1 + \mathfrak{s}_{\bf p}\right)  ~~,~~ \lambda_{2,{\bf p}} = q_{\bf p} \left( 1 - \mathfrak{s}_{\bf p}\right) ~~,
\end{eqnarray}
and
\begin{eqnarray}
    \tilde{\lambda}_{1,{\bf p}} &=&\tilde{q}_{\bf p} \left( 1 + \mathfrak{s}_{\bf p}\right)  ~~,~~
   \tilde{\lambda}_{2,{\bf p}}= \tilde{q}_{\bf p} \left( 1 - \mathfrak{s}_{\bf p}\right) ~~.
\end{eqnarray}
 
  {From the definition of the couplings, one can affirm that $0<\mathfrak{s}_{p}<1$. From another perspective, one can write down explicitly $\mathfrak{s}_{\bf p}$ as:}
\begin{equation}\label{ssol}
     \mathfrak{s}_{\bf p} = \frac{ \lambda_{1,{\bf p}}- \lambda_{2,{\bf p}}}{ \lambda_{1,{\bf p}}+ \lambda_{2,{\bf p}}} = \frac{ \tilde{\lambda}_{1,{\bf p}}- \tilde{\lambda}_{2,{\bf p}}}{ \tilde{\lambda}_{1,{\bf p}}+ \tilde{\lambda}_{2,{\bf p}}}~~.
\end{equation}
  {One can affirm from the equations above that $\mathfrak{s}_{\bf p}$ is fully determined by the coupling constants, and the choice made on Eq. \eqref{cases} imposes a strong link between the coupling constants. Mathematically, this link can be represented by the following identity:}
\begin{equation}
   {  \frac{{\lambda}_{2,{\bf p}}}{\lambda_{1,{\bf p}}} =  \frac{\tilde{\lambda}_{2,{\bf p}}}{\tilde{\lambda}_{1,{\bf p}}} }~~.
\end{equation}
Thus, going further in the analysis of this special case, one finds that the dissipative coefficients can be rewritten as follows:

\begin{equation}
  \xi^q_0 = \frac{\epsilon_{\bf p}^2}{2 m^2} \left( q_{\bf p}+ \tilde{q}_{\bf p}\right)~~,~~ \xi^q_i =\frac{\epsilon_{\bf p}}{2 m^2} {\bf p}_i\left( q_{\bf p}- \tilde{q}_{\bf p}\right)~~,~~  
\end{equation}
and
\begin{equation}
    \rho^q = \frac{\epsilon_{\bf p}}{2 m} \left( q_{\bf p}- \tilde{q}_{\bf p}\right)~~,
\end{equation}
One also has:
\begin{equation}
   \xi^d_\mu = 2\xi^q_\mu \mathfrak{s}_{\bf p}~~,~~   \rho^d = 2   \rho^q \mathfrak{s}_{\bf p}~~.
\end{equation}
 Now, the Retarded propagator can be written as follows:
\begin{equation}
     \check{\mathcal{S}}^R(p) = \frac{\slashed{p} + i \slashed{\xi}+m + i \rho^q}{\left(p_0 + i \xi^q_0\right)^2   -\epsilon_{\bf p}^2(1 + i \bar{\kappa} )^2 }~~,
\end{equation}
with $\bar{\kappa} =\frac{\epsilon_{\bf p}(q_{\bf p}-\tilde{q}_{\bf p})}{2 m^2}$, and its poles are given by:
\begin{equation}
     p^*_0=-\frac{i \epsilon_{\bf p}^2 (q_{\bf p}+\tilde{q}_{\bf p})}{2 m^2} \pm \sqrt{\left(\epsilon_{\bf p}+\frac{i \epsilon_{\bf p} (q_{\bf p}-\tilde{q}_{\bf p})}{2 m^2}\right)^2}~~,
\end{equation}
and after some algebraic manipulation, one gets:
\begin{equation}\label{poles1}
   {p_0^*}= \epsilon_{\bf p} - i \frac{\epsilon_{\bf p}^2}{m^2} q_{\bf p}~~,~~p_0^*=-\epsilon_{\bf p} - i \frac{\epsilon_{\bf p}^2}{m^2} \tilde{q}_{\bf p} ~~,
\end{equation}

 In the rest frame, it becomes:
\begin{equation}\label{restfr1}
     p_0^*({\bf p}=0) = m - i q_0 ~~\text{and}~~  p_0^*({\bf p}=0) - m -  i \tilde{q}_0~~.
\end{equation}

  {with $q_0 =\lim_{{\bf p} \to 0} q_{\bf p}$ and $\tilde{q}_0 =\lim_{{\bf p} \to 0} \tilde{q}_{\bf p}$}. Now, defining the energy of the particles by the real part of the poles, i.e., $ Re[p_0^*] = \epsilon_{\bf p} $ and the decay rate is expressed by the imaginary part $Im[p_0^*]=\Gamma_{\bf p} $. Therefore, by Eq. \eqref{poles1} it is easy to see that the dispersion relation obeys the relativistic dispersion relation. 

For the sake of stability, one needs to have $Im[p_0] = \Gamma_{\bf p}<0$, and one can see through the definition of $q_{\bf p}$ and $\tilde{q}_{\bf p}$ that all the poles respect this condition, and therefore one can affirm that the system is stable. 
Now, one can analyze one important property that can be extracted from the dispersion relation, the group velocity defined as $(v_g)_i=\frac{\partial \epsilon_{\bf p}}{\partial {\bf p}^i}$.   { It can be checked that the group velocity is given by $v_g = \frac{{\bf p} }{\epsilon_{\bf p}}$ and, therefore, is not modified by dissipation.}

\textbf{ Electromagnetic response:} It can be checked that the result of the Lyapunov equation shown in Eq. \eqref{specialcase} respects the generalized version of the fluctuation-dissipation theorem (FDT) for a fermionic open system given by 

\begin{equation}
\check{\mathcal{S}}^K(p) =   \left( \check{\mathcal{S}}^R(p) - \check{\mathcal{S}}^A(p) \right)  \mathfrak{s}_{\bf p}~~.
\end{equation}

By use of the Keldysh component of the propagator and using the fact that the Lyapunov solutions simplify to the on-shell scalar function $\mathfrak{s}_{\bf p}$, and since the rotation symmetry is maintained the spatial component of the current vanishes, and using Eq. \eqref{current}, the steady-state solution for the Lyapunov equation and the ansatz given by Eqs. \eqref{cases} one finds that the time component is the only non-null component of the current, and is written as follows:
\begin{widetext}
    \begin{eqnarray}\nonumber
          j^{st}_0 &=&    e \int_{-1}^1 d\lambda \int \frac{d^3 {\bf p}}{(2 \pi)^3}  \frac{  \lambda  (\gamma_--\gamma_+) \left((\gamma_-+\gamma_+)^2+4 \epsilon_{\bf p}^2\right)-  (\gamma_-+\gamma_+) \left(\gamma_-^2+4\gamma_- \gamma_++\gamma_+^2+4 \epsilon_{\bf p}^2\right)}{\sqrt{2}   \left((\lambda +1) \left(\gamma_+^2-2 (\lambda -1) \epsilon_{\bf p}^2\right)-\gamma_-^2 (\lambda -1)\right)^{3/2}}  \mathfrak{s}_{\bf p} ~~,\\
         &=&   e  \int \frac{d^3 {\bf p}}{(2 \pi)^3} g_{\bf p}\mathfrak{s}_{\bf p}~~.
    \end{eqnarray} 
\end{widetext}
  {where $\gamma_+ = \frac{\epsilon_{\bf p}^2}{m^2} q_{\bf p} $ and $\gamma_- = \frac{\epsilon_{\bf p}^2}{m^2} \tilde{q}_{\bf p} $. Therefore, the coupling with the bath can induce a charge imbalance in the system. Looking closer, the kernel $g_{\bf p}$ can be written as follows:}
\begin{equation}\label{gexpr}
    g_{\bf p} = -\frac{2 \left(8 m^4+3 \epsilon_{\bf p}^2 (q_{\bf p} +\tilde{q}_{\bf p})^2\right)}{4 m^4+\epsilon_{\bf p}^2 (q_{\bf p} +\tilde{q}_{\bf p})^2}  ~~,
\end{equation}
  {and in the dissipationless limit one finds $g_{\bf p} = -4$. In the ultra-relativistic limit, one finds $g_{\bf p} = - 6$. Interestingly, the rate between the dissipationless and the ultrarelativistic limits is $2/3$. Remarkably, the dissipation brings a relative weight to the charge imbalance, modifying the total charge imbalance depending on the strength of the dissipative couplings. }\\ 




\section{Final Remarks}
In this work, one seeks the formal description of a relativistic fermionic system that arose from the dynamics driven by the Lindblad master equation. The Lindblad equation ensures trace preservation, Hermiticity, and positive semi-definiteness of the reduced density matrix \cite{kamenev1}. Interestingly, the emerging Lagrangian is a non-Hermitian model that extends the Dirac action for relativistic fermions. The model is time-local, and the dissipative properties that bring non-locality in space coordinates depend on the coupling constants $\lambda_\sigma$. The resulting kinetic equation given by Eq. \eqref{kineq} generalizes the collisionless Boltzmann equation for the relativistic fermionic system by combining the spinorial structure of $\check{F}$ with the coupling to the dissipative terms $\check{Q}$ and $\check{D}$ and is our main result. By analyzing the stationary solutions of Eq. \eqref{kineq}, the problem simplifies, and one reaches the Lyapunov equation given by Eq. \eqref{stkin}. Thus, the solution of the Lyapunov equation given by Eqs. \eqref{Lyapsol1},\eqref{Lyapsol2},\eqref{Lyapsol3} are the most general solutions for systems that are independent of the chirality of the particles. More complex solutions that include spin and angular momentum contributions can be found if the chirality is considered.

 In our model some important differences require us to be cautious about drawing parallels with previous works on non-relativistic fermions (see, for instance, \cite{kamenev3}). The main distinction is that we do not employ the Nambu basis. This choice helps to compare our model with QFT theory but directly results in the property that the poles of the propagator do not occur in pairs of complex conjugates. This difference ultimately accounts for the asymmetry between particles/antiparticles. However, when analyzing the Keldysh component of the propagator, this asymmetry is not observed, since this component is constructed with both the retarded and advanced components of the propagator.

  {Especially looking at the application of the formalism shown in Section \ref{sec3}, where one assumes the cases where $n_{\bf p}= m_{\bf p}$, one finds that the complex poles of the retarded propagator do not bring instability for the particles nor antiparticles for any momentum $p$, and are a consequence of the Lindblad dynamical properties. Interestingly, if the particles (antiparticle) decouples from the Lindblad dynamics when the dissipative couplings $\lambda_{1,\sigma},\lambda_{2,\sigma} (\tilde{\lambda}_{1,\sigma},\tilde{\lambda}_{2,\sigma})$ are shut down. The resulting density matrix can be decomposed as $\rho_a \otimes \rho_b $ where the first obeys the von Neumann (GKSL) master equation whereas the second obeys the GKSL (von Neumann) master equation.  }

An interesting feature arose in our results: the apparent symmetry between particles and antiparticles in Eq. \eqref{H0} does not hold when looking at the poles of the propagator. In a heuristic way, one can see the problem as follows: the particle respects the equation of motion $(i \slashed{\partial} - m - i \slashed{\xi} - i \rho ) \psi = 0$, and doing a charge-parity transformation ($\mathcal{CP}$) one finds $\mathcal{CP} \psi \to \psi^c$ and $\psi^c$ respects $(i \slashed{\partial} + m + i \slashed{\xi} - i \rho ) \psi^c = 0$. One can identify these objects as $(S^R)^{-1}$ and $(S^A)^{-1}$, and they generate objects with different poles (complex conjugates of each other). But, due to causality and following the principle of stability, one analyzes only the poles of the retarded propagator which has $\Gamma_{\bf p} <0$. This condition is a reflex of the choice one made not to use the Nambu basis such as \cite{kamenev3}.  

%

The theorems about the uniqueness of the steady-state solution of the GKSL equation are based on the symmetries acting on the Lindblad operator, but with the assumption of the finiteness of the Hilbert space dimension (see for instance: Ref. \cite{Nigro:2018xov}). Since the quantum field theory demands an infinite dimensional Hilbert space, the uniqueness theorems are not applicable, and other attempts to deal with the symmetries have been made using Keldysh formalism (see for instance, ref. \cite{Altland:2020lbb}). 
When dissipation is strictly zero, the Hamiltonian, $H_f$ fully determines the unitary time evolution. 
However, for any non-zero dissipation strength, the system will reach a steady state,  to which it will tend at sufficiently large times. In our pedagogical model, one can affirm that the steady state is determined by $\mathfrak{s}_{\bf p}$, which in turn is determined by the coupling constants $\lambda_i,\tilde{\lambda}_i$ as shown in Eq. \eqref{ssol}.   { One also finds that the condition $0< \mathfrak{s}_{\bf p}<1$ implies that the scalar Wigner distribution can be interpreted by an effective number occupation.  An interesting feature that can be analyzed is to go beyond the assumptions given by Eq. \eqref{cases}. By doing that, the vector and tensor components of the Wigner distribution do not vanish, and the hydrodynamics of the system can be a target of further investigation.} 

It is essential to highlight that in this work, we find a generalization version of the Boltzmann equation for a relativistic open fermionic system (see Eq. \eqref{kineq}).
Using the Keldysh formalism combined with the spinor properties, one has successfully determined the Lyapunov equations for stationary Wigner distributions that generalize equilibrium kinetic distributions using the Dyson-Schwinger equations. In Section \ref{sec3} one recovers the FDT solution for a non-Hermitian relativistic fermionic system.

In a context where an introduction of a chemical potential makes no sense, the stable fermionic solution of the Lyapunov equations results in a nontrivial stationary total charge imbalance, which can be calculated based on the coupling constants between the system and the bath. Although a non-Hermitian model presented in Section \ref{sec3} is limited due to the simplistic assumptions, it can be a first approximation to a new way of looking at the matter/antimatter imbalance problem. Because of the memory-less character typical of Markovian dynamics, the dependence on the initial conditions does not appear in the model presented. Thus, besides our initial results being just a toy model that generates charge imbalance due to dissipation, a more realistic realization of the model could be a way to describe the particle/antiparticle asymmetry in the universe successfully. Due to the memoryless property of the Lindblad formalism, the asymmetry could be explained by the proper dynamics of the system instead of a problem of some cosmological initial conditions. This effect could also be important in dynamic symmetry-breaking phenomena, modifying the critical density where the symmetry is restored.

Furthermore, suppose one generalizes the couplings to the bath to include the chirality of the particles. In that case, one can find the non-Hermitian terms $\bar{\psi} \gamma_5 \psi$ and $\bar{\psi}\gamma_5 \gamma_\mu \psi$, recovering some results found in the recent literature (for instance, see Refs. \cite{PT1,PT2,PT3}. Another way to describe more general systems is to introduce extra terms to the Hamiltonian, as $(\hat{a}_{\sigma}\hat{a}_{\sigma'} + h.c.)$ terms. In condensed matter, this term generates the so-called quantum heating effects \cite{qr1,qr2} and can bring exciting modifications to the fermionic system described in this work. The generalization of coupling with the bath to a nonlinear version can also generate new phenomena. 
Regardless of the initial state, quadratic Lindblad systems with linear jump operators could sometimes contain instabilities, resulting in an exponential runaway behavior. Despite our model being stable, more realistic models can suffer from instabilities and it is well-known that nonlinearities in the system can eventually curtail this instability. Nonlinearities rule this aspect of the system's behavior, as Nambu-Jona-Lasinio models \cite{gomes21,Nambu61A,Nambu61B}, and can be an interesting line of research.

\section{Appendix}\label{appA}
This appendix shows the orthonormalized Dirac eigenvectors used in the article. Starting from the Dirac equation in four dimensions, one has:
\begin{equation}
    (i \gamma^\mu \partial_\mu - m ) \psi(x) = 0
\end{equation}
with $\psi(x)$ a generic four-component spinor and $\gamma^\mu$, $\mu=0,i$ ($ i=x,y,z$) the $4 \times 4$ Dirac matrices in the Dirac representation can be written as:
\begin{equation}
   \gamma^0 = \begin{pmatrix}
        \mathds{1}_{2 \times 2} & 0 \\
        0 & -\mathds{1}_{2 \times 2} 
    \end{pmatrix} ~~,~~ \gamma^i = \begin{pmatrix}
        0 & \sigma^i \\
        - \sigma^i & 0
    \end{pmatrix}~~.
\end{equation}
where $\sigma^i$ are the Pauli matrices. The plane-wave solutions are eigenvectors representing the particle given by:
\begin{equation}
    u_{{\bf p},1} = N\begin{pmatrix}
    1\\
    0 \\
    \frac{p_z}{\epsilon_{\bf p} + m}\\
    \frac{p_+}{\epsilon_{\bf p} + m}
            \end{pmatrix}~~,~~
            u_{{\bf p},2} = N\begin{pmatrix}
    0\\
    1 \\
    \frac{p_-}{\epsilon_{\bf p} + m}\\
    -\frac{p_z}{\epsilon_{\bf p} + m}
            \end{pmatrix}
\end{equation}
The eigenvectors representing the antiparticle are given by:
\begin{equation}
    v_{{\bf p},1} = N\begin{pmatrix}
    \frac{p_-}{\epsilon_{\bf p} + m}\\
    -\frac{p_z}{\epsilon_{\bf p} + m}\\
    0\\
    1
            \end{pmatrix}~~,~~
            v_{{\bf p},2} = N\begin{pmatrix}
             \frac{p_z}{\epsilon_{\bf p} + m}\\
    -\frac{p_+}{\epsilon_{\bf p} + m}\\
    1\\
    0 \\
            \end{pmatrix}
\end{equation}
with $p_\pm = p_x \pm i p_y$, $N=\sqrt{\epsilon_{\bf p}+ m}$ such way $u^\dagger u= v^\dagger v = 2 \epsilon_{\bf p}$. The eigenvectors respect the following identities:
\begin{equation}
\bar{u}_{{\bf p}, s} u_{{\bf p}, s'} =2 m \delta_{s s'}~~,~~ \bar{v}_{{\bf p}, s} v_{{\bf p}, s'} =-2 m \delta_{s s'}
\end{equation}
and
\begin{equation}
\bar{u}_{{\bf p}, s} \gamma^\mu u_{{\bf p}, s'}=\bar{v}_{{\bf p}, s} \gamma^\mu v_{{\bf p}, s'} =2 \hat{p}^\mu \delta_{s s'}~~,~~\hat{p}^\mu = (\epsilon_{\bf p}, {\bf p})~~.
\end{equation}
The eigenvectors also obey the completeness relation given by:
\begin{equation}
   \sum_s u_{{\bf p}, s} \bar{u}_{{\bf p}, s} = \hat{\slashed{p}} + m ~~,~~ \sum_s v_{{\bf p}, s} \bar{v}_{{\bf p}, s} = \hat{\slashed{p}} - m~~.
\end{equation}
One can write the particle/antiparticle projectors given by:
\begin{equation}
    \mathbb{P} = \frac{\hat{\slashed{p}}+ m}{2 m} ~~, \mathbb{P}^c = \frac{-\hat{\slashed{p}}+m}{2m} ~~, 
\end{equation}
such way, one can write:
\begin{equation}
     \mathbb{P} u_{{\bf p}, s} = u_{{\bf p}, s}~~,~~   \mathbb{P}^c u_{{\bf p}, s} = 0~~,~~ \mathbb{P}^c v_{{\bf p}, s} = v_{{\bf p}, s}~~,~~   \mathbb{P} v_{{\bf p}, s} = 0
\end{equation}

\section*{Acknowledgments}
The author is grateful to MFSA for major revisions. YMPG is supported by a postdoctoral grant from  {}Funda\c{c}\~ao
Carlos Chagas Filho de Amparo \`a Pesquisa do Estado do Rio de Janeiro
(FAPERJ), grant No. E26/201.937/2020. 
%


%

\end{document}